# ABOUT THE RECALIBRATION OF THE SUNSPOT RECORD

Georgieva K.[1], Kilçik A.[2], Nagovitsyn Yu.A.[3,4], Kirov B.[1]

[1]*SRTI-BAS, Sofia, Bulgaria*
[2]*Akdeniz University, Antalya, Turkey*
[3]*Central Astronomical Observatory at Pulkovo, St. Petersburg, Russia*
[4]*St. Petersburg State University of Aerospace Instrumentation, St. Petersburg, Russia*

## О ПЕРЕКАЛИБРОВКЕ ЧИСЛА СОЛНЕЧНЫХ ПЯТЕН

Георгиева К.[1], Килчик А.[2], Наговицын Ю.А.[3], Киров Б.[1]

[1]*ИКИТ-БАН, София, Болгария*
[2]*Средиземноморский университет, Анталия, Турция*
[3]*ГАО РАН, Санкт Петербург, Россия*
[4]*ГУАП, Санкт Петербург, Россия*

*В 2015 было прекращено вычисление Международного (Цюрихского) числа солнечных пятен, и его заменили новой „перекалиброванной" серией. Был предложен и новый ряд Числа групп солнечных пятен. В результате эти две серии, которые в оригинале отличались в основном по долгосрочным трендам, были приведены „в согласие", и у обоих тренд был минимизирован. Эта перемена привела к новым реконструкциям длительных изменений солнечной радиации и к пересмотру влияния вариаций солнечной активности на изменения климата. Мы использовали данные о числе солнечных пятен и групп солнечных пятен обсерватории с непрерывными и однородными наблюдениями с тем, чтобы оценить целесообразность этой перемены.*

The number of sunspots is the longest instrumental data record of solar activity, and is widely used for studies of both the Sun, and the solar effects on the terrestrial system. It plays a critical role in evaluating the relative contribution of natural versus anthropogenic factors for the observed climatic changes because it is the proxy for reconstructing the variations of total and spectral solar irradiance which are key parameters in climate models.

Total solar irradiance has been measured since 1978 [1]. For earlier periods, the evaluation of irradiance variations is based on proxies. The irradiance is determined by the darkening due to sunspots, plus brightening due to ephemeral regions, network, and faculae, plus the contribution of the quiet Sun [2]. The sunspots' darkness and the faculae's brightness are determined by their magnetic field [3] which is found to be proportional to their areas [4]. Further, there is a correlation between the total areas of faculae and sunspots [5], and the total area of sunspots is in turn proportional to the number of sunspots [6]. Full disc magnetograms from which sunspots' darkness and faculae's brightness can be estimated are available since 1974, sunspot and facular areas as a proxy for the magnetic field since 1874, sunspot number as a proxy for sunspot areas since 1700, and sunspot groups as a proxy for sunspot number since 1610. Therefore, progressively longer reconstructions of solar irradiance include progressively fewer directly measured parameters, and all reconstructions starting before 1874 are based on the number of sunspots or sunspot groups as the only instrumentally measured parameter.

Until very recently there were two sunspot series used in reconstructions of long-term variations of the solar irradiance and other solar activity parameters:

- The original "relative sunspot number", known also as "Zurich international sunspot number" or "Wolf number", WN, was defined by Wolf as

$$WN = k\,(10 \times GN + SN), \qquad (1)$$

where GN is the number of sunspot groups, and SN – the number of individual sunspots [7]. The scaling coefficient k accounts for the differences in measurement technique, viewing conditions, observer's experience, etc. This data series covers the period from 1700 to 2015.

- A second series expanded the sunspot record back to 1610: the group sunspot number WG [8]. It is based on the parameter GN in equation (1) – the number of sunspot groups which is more reliably determined and allows the inclusion of earlier observations. The normalization factor scaling WG to WN values for the period of 1874–1976 is 12.08:

$$WG = 12.08 \times GN. \qquad (2)$$

Some reconstructions of solar irradiance based on the sunspot record, and studies using them to evaluate solar influences on climate, used WN as a proxy, others WG, and still others – both, because WN is not available before 1700, and WG is not available after 1995 [9]. The main difference between the two series is the long-term trend which is much larger in the case of WG (Fig. 1). Consequently, some studies using WG come to the conclusion that the second half of the 20$^{th}$ century was a period of unusually high solar activity, and that there was a substantial increase in the solar irradiance and overall solar forcing since the Maunder minimum, leading to estimated much higher role of the Sun in climate changes, as compared to studies using WN.

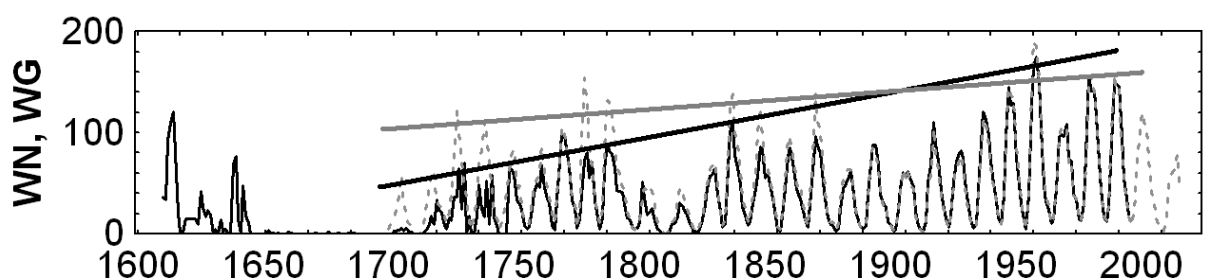

**Fig. 1.** Group sunspot number (black solid line) and Wolf sunspot number (grey dotted curve), yearly averages, with the respective trends.

The Wolf number was provided by the Zurich Observatory until 1980, after which its production was transferred to the Royal Observatory of Belgium in Brussels [10] where the "Sunspot Index Data Center" (SIDC) as a part of the World Data Center was established. SIDC existed until 2013, when it was replaced by the new "Sunspot Index and Long-term Solar Observations" (SILSO) data center. This was not a simple transition, but, as stated by SILSO's director Frederic Clette [http://www.sidc.be/silso/news001], it was "just the visible part of an ongoing modernization and expansion of the historical Sunspot Number series". Actually, the process of replacement of the original sunspot data by an entirely new data series started in 2011 with the first Sunspot Number (SSN) Workshop, followed by several more similar workshops during 2012–2014. The justification of all this activity was that, "given the importance of the reconstructed time series, the co-existance of two conflicting series is a highly unsatisfactory solution which should be actively addressed" [11]. The stated goals were to "rectify discrepancy between Group and International sunspot number series", and to publish "a vetted and agreed upon single sunspot number time series" [11, 12]. As a result of this series of workshops and the related activities, a new sunspot number was produced [12]. In the meantime, the Group Sunspot Number data series WG was also reconstructed [13]. In Fig.2, presenting the comparison between these two new series, two important features are seen: the two series match very closely, and both have practically no long-term trend. At a press briefing during the IAU XXIX General Assembly it was announced that the corrected sunspot history suggests that "rising global temperatures since the industrial revolution cannot be attributed to increased solar activity" (https://www.iau.org/news/pressreleases/detail/iau1508/).

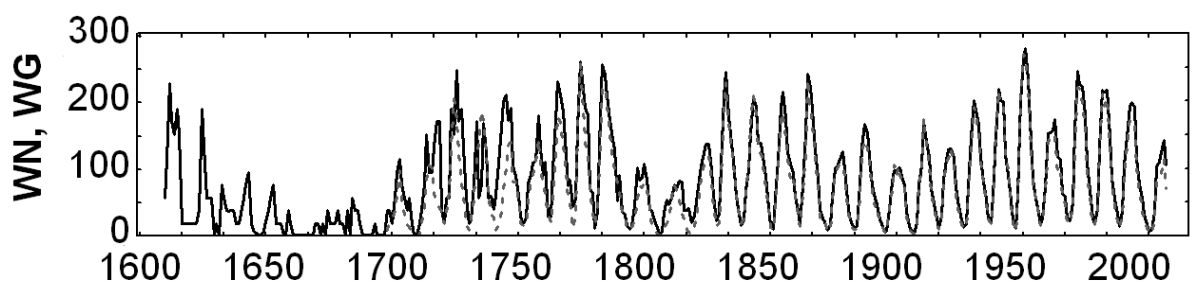

**Fig. 2.** The new versions of the group sunspot number (black solid line) and Wolf sunspot number (grey dotted curve), yearly averages.

Following the IAU XXIX General Assembly, on July 1, 2015 WDC-SILSO stopped the production of the original sunspot data and replaced it by the new ones (http://www.sidc.be/silso/datafiles). However, the goal to publish an "agreed upon single sunspot number time series" was not achieved, neither for the sunspot number nor for the sunspot group number. On the contrary, the new series were not unanimously accepted, and led to the ongoing creation of more still newer alternative series [http://www.spaceclimate.fi/SC6/presentations/session2b/Frederic_Clette_SC6.pdf].

The other goal – to reconcile the two series – was more than fulfilled as not only the differences between the new WN and WG were minimized, but also the trend in both series was removed. In the course of the work towards this goal, some questions were raised regarding the homogeneity of each of the original time series [11], and the new ones claim to have answered them. In the present study we are not dealing with the possible flaws in the original WN and WG, neither with the applied corrections to remove these flaws. Instead, we are examining whether it is justified to set as a goal to minimize the discrepancies between the two series.

Our earlier studies [14, 15] have demonstrated that the relations between the sunspot's magnetic field and area, between the total areas of sunspots and faculae, and between the sunspots' total area and number, all vary from cycle to cycle. The long-term variations of these relations can give some information about the long-term variations of the solar dynamo, and should be taken into account in the reconstructions. We will now compare the ratio between sunspot number and sunspot groups in different cycles.

The National Geophysical Data Center (ftp://ftp.ngdc.noaa.gov/STP/ SOLAR_DATA) provides various active region parameters including the sunspot group (SG) classification and sunspot count – the number of sunspots on the solar disk (SSC) determined for each day. The database collected by the US Air Force/Mount Wilson Observatory includes measurements from Learmonth, Holloman, and San Vito Solar Observatories. We use the Learmonth data (LEAR) as the principal data source, while the data gaps are filled in with observational records from one of the other stations, so that a nearly continuous time series is produced.

Using data for the four most recent solar cycles, Kilcik et al. [16] separated SGs into large and small groups and found that, in general, the large SGs peaked about 2 yr after the peak of the small SG and that the temporal variations of the large SGs better correlate with the solar and geomagnetic activity indices. Later, Kilcik et al. [17] modified the separation criteria by additionally taking into account the sunspot evolution, thus separating active regions into four types: (1) "S" – simple (Zurich classes A and B), (2) "M" – medium (class C), (3) "L" – large (classes D–E and F), and (4) "F" – final (class H) types. Here we use this new separation scheme to investigate temporal variations of the SSCs and the number of SGs. The data includes the time interval from January 1982 to December 2015: the descending branch of cycle 21 (1982–1986), cycles 22 (1986–1996) and 23 (1996–2008), and the first half of cycle 24. To mitigate the effect of data gaps, we derived the total daily number of SGs and SSCs of a given group type and then averaged it over the month, thus obtaining a parameter, essentially independent of data gaps. To remove short-term fluctuations and to reveal long-term trends, the monthly averaged time series were smoothed with 12-month moving average.

Fig. 3a presents the total sunspot count and the total number of sunspot groups as measured by LEAR in the last 4 cycles. Though their variations are very similar as expected, there is no one to one correspondence between them.

To illustrate this further, in Fig. 3b the average number of sunspots per group are plotted with SSC added for reference. Clearly seen is that this ratio is not constant: it has a strong solar cycle variation, and varies from cycle to cycle. The most obvious feature is the deep minimum between cycles 23 and 24, and the lower maximum in the maximum of cycle 24. That is, not only the number of sunspots and the number of groups are less in cycle 24 and the minimum before it than in the previous cycles and minima, but also the average number of sunspots in the groups is much lower. In Fig. 3c the time evolution of group types is compared. Though the total SSC and SG in cycles 22 and 23 are very similar, the distribution of the different group types is very different. L groups are more than S groups in cycle 23 than in cycle 22 maximum, while the opposite is true for the S groups. In both maxima the number of L groups is higher than that of S groups, while their number is almost equal in cycle 24, due to the substantial decrease of L groups and a small increase of S groups. The number of M groups decreases slowly but persistently in consecutive maxima. The number of F groups is equal in the maxima of cycles 22 and 24 and higher in cycle 23 consis-tent with the higher number of L groups. Variable is also the number of sunspot per group in the different group types (Fig. 3d). This number is a little lower in cycle 23 than in 22, and much lower in cycle 24 for L groups, persistently decreases from cycle to cycle for M groups, and practically doesn't change for F groups. Interesting features are the strong decrease of the number of sunspots in S groups in the maximum of cycle 23, and the prevalence of M groups throughout almost the whole cycle 24.

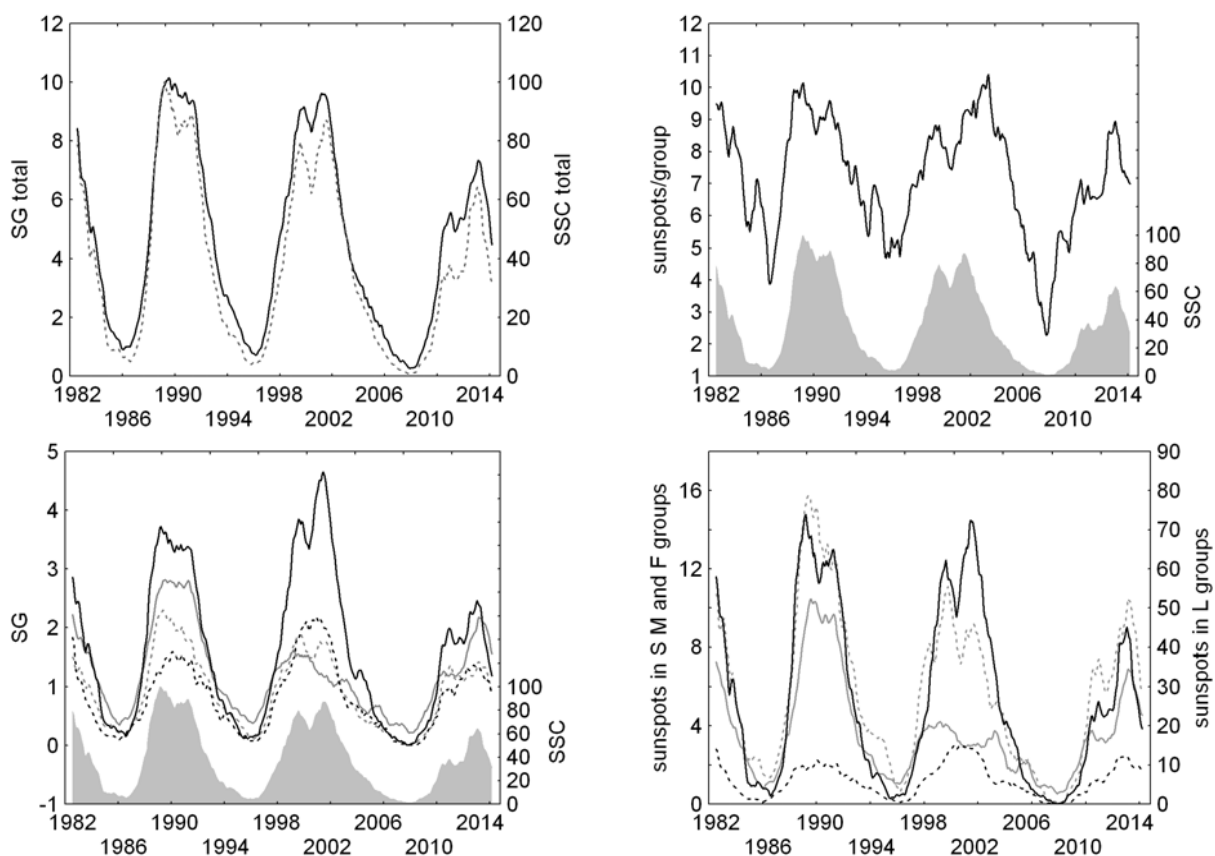

**Fig. 3** (**a**) total number of sunspots (grey dotted line) and total number of sunspot groups (black solid line) measured in LEAR; (**b**) average number of sunspots per group (black solid line) and the total sunspot count (grey area); (**c**) number of SG of type S (grey solid line), M (grey dotted line), L (black solid line) and F (black dotted line); line); (**d**) average number of sunspots in groups of type S (grey solid line), M (grey dotted line), L (black solid line) and F (black dotted line)

The presented results based on data from a single observatory with continuous and homogenous observations demonstrate that the variation in the ratio between the sunspot counts and number of sunspot groups, respectively between the sunspot number and group sunspot number calculated from them in the way in which the original WN and WG are calculated, is a real feature, and not a result of changing observational instruments, observers' experience, calculation schemes, etc. Therefore, the attempts to "rectify discrepancy" between the two data series and to "reconcile" them are not justified and the resulting new "recalibrated" series are misleading. Moreover, in this way important information is lost which can shed light on the long-term evolution of the Sun and the solar dynamo.